\documentclass[preprint]{aastex}
\usepackage{natbib}
\usepackage[caption=false]{subfig}
\usepackage{graphicx}
\usepackage{acronym}
\usepackage[breaklinks=true]{hyperref}
\usepackage{bm}
\usepackage{epstopdf}
\usepackage{multirow}
\usepackage{gensymb}
\usepackage{afterpage}
\begin{document}
\title{Image-Optimized Coronal Magnetic Field Models}
\author{Shaela I.\ Jones\altaffilmark{1}\email{shaela.i.jones-mecholsky@nasa.gov} \and Vadim Uritsky \altaffilmark{1} \and Joseph M.\ Davila}
\affil{NASA Goddard Space Flight Center, Code 670, Greenbelt, MD  20771}
\altaffiltext{1}{Catholic University of America}
\email{shaela.i.jonesmecholsky@nasa.gov}
\begin{abstract}
We have reported previously on a new method we are developing for using image-based information to improve global coronal magnetic field models.  In that work we presented early tests of the method which proved its capability to improve global models based on flawed synoptic magnetograms, given excellent constraints on the field in the model volume.  In this follow-up paper we present the results of similar tests given field constraints of a nature that could realistically be obtained from quality white-light coronagraph images of the lower corona.  We pay particular attention to difficulties associated with the line-of-sight projection of features outside of the assumed coronagraph image plane, and the effect on the outcome of the optimization of errors in localization of constraints.  We find that substantial improvement in the model field can be achieved with this type of constraints, even when magnetic features in the images are located outside of the image plane.
\end{abstract}
\acrodef{pfss}[PFSS]{Potential Field Source Surface}
\acrodef{euvi}[EUVI]{Extreme Ultraviolet Imager}
\acrodef{pos}[POS]{plane of the sky}
\acrodef{pb}[pB]{polarization brightness}
\acrodef{stereo}[STEREO]{Solar TErrestrial RElations Observatory}
\acrodef{cme}[CME]{coronal mass ejection}
\acrodef{3d}[3D]{three-dimensional}
\acrodef{los}[LOS]{line-of-sight}
\acrodef{imf}[IMF]{interplanetary magnetic field}
\acrodef{nlfff}[NLFFF]{Non-Linear Force Free Field}
\acrodef{lfff}[LFFF]{Linear Force Free Field}
\acrodef{rms}[RMS]{root-mean-square}
\acrodef{euv}[EUV]{extreme ultraviolet}
\acrodef{mae}[MAE]{mean absolute error}
\acrodef{sht}[SHT]{spherical harmonic transform}
\acrodef{gong}[GONG]{Global Oscillation Network Group}
\acrodef{mhd}[MHD]{magnetohydrodynamics}
\acrodef{spp}[SPP]{Solar Probe Plus}
\acrodef{so}[SO]{Solar Orbiter}
\acrodef{mlso}[MLSO]{Mauna Loa Solar Observatory}
\acrodef{lasco}[LASCO]{Large Angle and Spectrometric Coronagraph Experiment}
\acrodef{fom}[FOM]{figure of merit}
\section{Introduction}
Due to the long-standing difficulty of measuring the magnetic field of the solar corona, a series of approaches to extrapolating the field from photospheric measurements have been devised and implemented over the course of several decades.  \citet{mackay12} and \citet{regnier13} provide excellent reviews of the current most popular methods, including \ac{pfss}, \ac{nlfff} and \ac{mhd} models.  As those authors have described, each of the above methods has positive and negative aspects.  The \ac{pfss} models \citep{altschuler69,schatten69} neglect currents in the corona; on the other hand they are quick to compute and produce a unique solution for a given set of boundary conditions.  \ac{nlfff} models are more physically realistic in that they incorporate field-aligned currents, but their form can be quite sensitive to the treatment of the problem and of the boundary conditions \citep{schrijver06, metcalf08, derosa09}.  Coronal \ac{mhd} models \citep{riley11} offer much more inclusive physics, temporal evolution, and full characterization of the plasma, but due to their computational complexity, often require users to make trade-offs between resolution, computation time, and fidelity to the physics.

A difficulty faced when using any of the above methods is the limited accuracy of the photospheric measurements used as boundary conditions.  There are a number of error sources that negatively affect photospheric magnetograms: unobserved polar regions, saturated absorption lines, unresolved features, unfavorable perspective for regions near the limb/poles, the finite thickness of the photospheric layer, etc.  Synoptic magnetograms, which are built from combinations of disk magnetograms measured over a solar rotation, are necessary for creating global models, but suffer from the additional complication that the field has evolved over the time period of their formation.   \citet{riley14} conducted a systematic comparison of synoptic magnetograms from six different observatories and found substantial and irregular differences between them, concluding that they could find no ``ground truth" about the magnetic field of the photosphere.
\begin{figure}[ht!]
  \centering
\includegraphics[width=8.5cm]{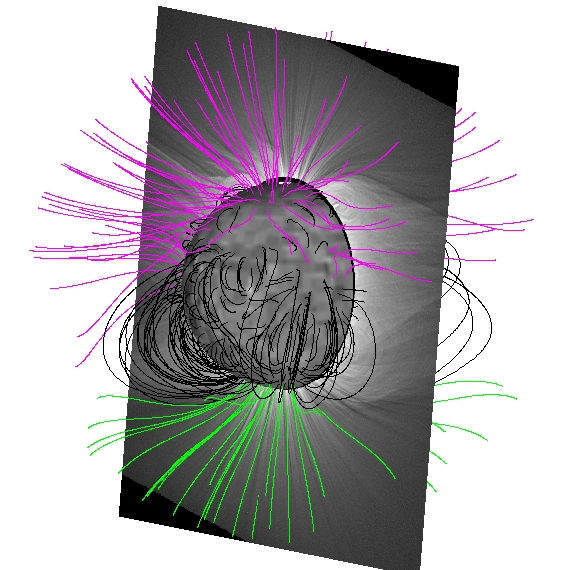}
 \caption{Intersection of a solar eclipse image of the corona and a \ac{pfss} coronal magnetic field model.  Solar eclipse image created by Miloslav Druckm{\"u}ller, Peter Aniol, and Jan Sl{\'a}de{\u c}ek, as featured in \citet{druckmuller14}.  URL: \url{www.zam.fme.vutbr.cz/~druck/eclipse/Ecl2008m/Tse2008_500_mo1.png}}
\label{fig:demo_figure}
\end{figure}

In \citet{jones16}, hereafter Paper I, we described a technique we have devised for creating global coronal magnetic field models that are less dependent on the accuracy of the underlying synoptic magnetogram, by incorporating morphological information drawn from coronal images.  It often happens that coronal images show quasi-linear features or boundaries that appear to delineate the local magnetic field.  We can quantify the orientation of these features at discrete locations $\textbf{r}$ using an azimuth angle $\theta_o$ measured counterclockwise from the horizontal.  Figure \ref{fig:demo_figure} shows the intersection of a solar eclipse image plane and a magnetic field model.  Features seen in the image plane delineate the apparent direction of the magnetic field, and the measured orientation angles (hereafter referred to as constraints) quantify this at discrete locations.  As the \textbf{Sun} rotates over the course of about two weeks, angles from several such planes can be combined to constrain the field throughout the volume.  The discrepancy between the constraints, $\{\theta_o\}$, and the orientation of the magnetic field at the same locations in a global coronal model, $\{\theta_m\}$, can be quantified using the objective function:
\begin{equation}
\label{eqn:objective}
J=\beta \sum_{k=1}^N (\theta_{o,k} - \theta_{m,k})^2+\gamma\,  O
\end{equation}
In this function, $\beta$ and $\gamma$ are normalization constants, the sum runs over all $N$ identifiable constraints, and $O$ represents additional optional terms that can be included based on \emph{a priori} notions about the solution.  From inspection of equation \ref{eqn:objective}, one can see that the greater the discrepancy between the model and the image-based constraints, the greater the value of $J$.  Using this method for quantifying the discrepancy, we can discriminate between different coronal magnetic field models.  

In Paper I we described a method to use this discriminatory power to compensate for uncertainty in the synoptic magnetogram flux values used to extrapolate global \ac{pfss} models.  Specifically, our method uses the Downhill Simplex optimization method of \citet{nelder65} to perturb the synoptic flux values until optimal agreement is found between the model and the image-based constraints, via a minimum of the function $J$.  As described in Paper I, we found it more efficient to optimize the spherical harmonic coefficients of the synoptic magnetogram rather than altering the magnetogram flux values themselves; this allowed us to make global changes to the magnetogram while optimizing a relatively small number of lowest-order spherical harmonic coefficients.

Optimization techniques have been used to compare coronal images with models in the past.  Aschwanden and co-authors \citep{aschwanden13b,aschwanden12,aschwanden10}, used a set of mathematical magnetic monopoles buried beneath a surface to approximately duplicate magnetogram measurements beneath an active region.  They then created a \ac{nlfff} model for the active region based on their approximate magnetogram, determining the amount of twist required on each field line by optimizing agreement between the model and coronal loops whose positions were determined from \ac{euv} images.  \citet{conlonandgallagher10} constrained active region field lines in a \ac{lfff} model using \ac{euv} observations. \citet{malanushenko09} used bright field lines observed in \ac{euv} and X-ray images to constrain a \ac{lfff} model, finding the most consistent values for the height and twist of the observed coronal loops.  In later papers \citep{malanushenko12, malanushenko14} the authors used these fitted values as constraints to iteratively fill a \ac{nlfff} model.  This last method has the advantage that it is not dependent on photospheric vector magnetograms as boundary conditions.  However, all of these methods that make use of coronal loop measurements are limited in application to regions where coronal emission is strong (\emph{e.g.} active regions), and require a specialized method for identifying the locations of the coronal loops.

In this paper we will describe the advantages of working with white-light coronagraph images for optimization of global coronal magnetic field models.  We will also address a common concern expressed by members of the heliophysics community when discussing this method: namely, \ac{los} projection.  Finally, we will present further testing of our approach, specifically testing with the kind of constraint sets $\{\theta_o\}$ that could realistically be obtained from coronagraph images. 
\section{Advantages of Coronagraphs for Global Field Model Optimization}
Working with white-light coronagraph images for optimizing global coronal magnetic field models offers several advantages, particularly for the purposes of bounding heliospheric models.  One is that coronagraph images show the corona at higher solar altitude, offering data more suited to the determination of the open/closed magnetic field boundary.  This advantage is particularly important given the diminished reliability of magnetograms near the polar regions; the polar field strength has a significant effect on the configuration of the open/closed field boundary.  Another way of looking at this is that higher altitude observations are dominated by the large-scale spherical harmonics of the magnetic field, which evolve more slowly \citep{altschuler77} and are increasingly important farther out in the heliosphere.

Additionally, the large-scale nature of the features observed in coronagraphs means that the orientation of the magnetic field in the model generally does not vary greatly with small changes in position.  When looking at small-scale features like coronal loops, a small error in the observer's determination of the loop position can mean a very big difference in the orientation of the model magnetic field.  In contrast, constraints based on coronagraph images are typically more forgiving, an important issue which will be discussed further in section \ref{section:los}.

It's important to note, too, that while our focus has been primarily in improvement of the global scale coronal field model, this improvement may also be beneficial to our understanding of smaller-scale coronal components.  \citet{schrijver06} found that provision of accurate boundary conditions on all sides of the model region greatly improved the accuracy of \ac{nlfff} models of active regions; these boundary conditions are often set via the calculation of a global \ac{pfss} model, as are the initial states of \ac{mhd} models.
\section{Line-of-Sight Projection and Localizing Image Features in Three Dimensions}
\label{section:los}
The remark above about the forgiving nature of the large-scale coronal structures brings up an objection that is sometimes raised when this method is presented: \acf{los} confusion.  Because the corona is optically thin, the intensity measured in a particular image pixel is actually an integral of the scattered photospheric light from electrons all along that pixel's \ac{los}.  Given this fact it is unavoidable (and one often observes in, for example, the \ac{lasco} images) that coronal features located outside of the coronagraph image plane appear in the image in projection.  In movies made from these images, a large streamer rotating around the Sun might be seen crossing the image from left to right and back again as it rotates in front of and then behind the Sun.  Given this ambiguity in the third dimension, how can we accurately place our coronagraph-based constraints in \ac{3d} space?

We start by extracting constraints from \ac{pb}, rather than total brightness, coronagraph images.  Due to the scattering geometry and the fact that light is polarized perpendicular to its direction of travel, photospheric light scattered from electrons in the image plane is more polarized than light scattered from those away from it \cite{billings66,howard09}.   The \ac{pb} images, produced by combining coronagraph images in which a polarizing filter has been used to measure the linear polarization in several different directions, place a much greater emphasis on the material near the image plane, or \ac{pos}.  To compare features seen in \ac{pb} coronagraph images with a \ac{3d} model, we assume that the features are located precisely in the \ac{pos}, and calculate the \ac{3d} position of the features using the known position of the observer and the location of the feature in the image.  

This assumption will naturally result in some error in the locations $\textbf{r}$ associated with the coronagraph-based constraints; however, the spatial extent of the features we are trying to optimize using this method is large compared to the expected error in the constraint location.  In order to maintain a reasonable number of parameters to be optimized, we optimize only the coefficients of spherical harmonics with $l \leq 6$ - all other coefficients are held constant.  In this case the minimum distance between longitudinal nodes of the optimized harmonics is $15\degree$.  

The impact of errors due to \ac{los} projection is additionally limited by the statistical nature of the optimization.  In searching for the optimal set of harmonic coefficients, the optimization routine does not need to find a set that creates a perfect match between the images and the model, rather one that best satisfies the set of available constraints.   In section \ref{section:los_test} we present a theoretical test of the method that includes error in the locations of the hypothetical image-based constraints and discuss the impact of these errors on the optimization result.
\section{Testing with Coronagraph-Like Constraints}
In Paper I we created a test problem for our magnetic field model optimization method using a \ac{gong} magnetogram.  We used the magnetogram to create a \ac{pfss} model (the ``ideal" field) and  sampled the orientation of the magnetic field in twelve hypothetical image planes.  We then perturbed four spherical harmonic transform coefficients of the \ac{gong} magnetogram.  We fed this perturbed magnetogram to the optimization software in place of the proper one, asking it to find a \ac{pfss} model, starting with the perturbed magnetogram, that best matched the ideal field orientations.  The optimization software performed very well on this test, correcting the perturbed coefficients without introducing significant error to the correct ones.

A logical next step following this test is to construct theoretical tests with constraint sets $\theta_o$ that could realistically be obtained from coronagraph images.  In Paper I the optimization software was provided the orientation of the field in every voxel in twelve image planes, without error.  In the following tests the constraint locations are distributed near the equator and are clustered in radius near a hypothetical coronagraph occulter.  The latitudinal locations of the constraints are drawn from a normal distribution with a mean of zero (meaning centered on the equator) and standard deviation of $20\degree$.  The radial locations are drawn from a distribution with a minimum of $1.05 R_{\sun}$, a maximum of $2.5 R_{\sun}$, and which peaks near $1.1 R_{\sun}$.  Subsequently the coordinates were adjusted to account for a non-zero solar-B angle, with values drifting from $1.2\degree$ to $2.8\degree$ over the set of image planes to simulate the perspective of an observer on the Sun-Earth line.  We sampled the image-plane component of the ideal field model in $14$ planes to collect a total of $4300$ constraints (just over 300 per image plane).  The sampling locations on each plane were chosen randomly, without regard for the resolution of the model, such that some points were likely sampled (by interpolation) at locations within the same voxel, and may represent much the same information.  
\subsection{Errors in Orientation of Observed Features}
\label{section:noisy_constraints}

In addition to being distributed primarily near the hypothetical streamer belt, and primarily at altitudes near an occulter edge, coronagraph-based constraints would have some degree of error in the measured orientation angles.  To examine the effect of this error on our optimization results, we added normally distributed errors to the field orientations $\{\theta_o\}$ sampled from the ideal field.  We then fed them to the optimization routine along with the perturbed transform, as was done in Paper I.  This test was repeated for error distributions with standard deviations of $2\degree$, $5\degree$, $10\degree$, $15\degree$, and $20\degree$.  Equation \ref{eqn:objective} assumes error-free constraints $\theta_{o}$, so as we increase the error in the measurements we can expect that the agreement between the optimized and ideal fields will deteriorate.
\begin{figure}
  \centering
\includegraphics[width=11cm]{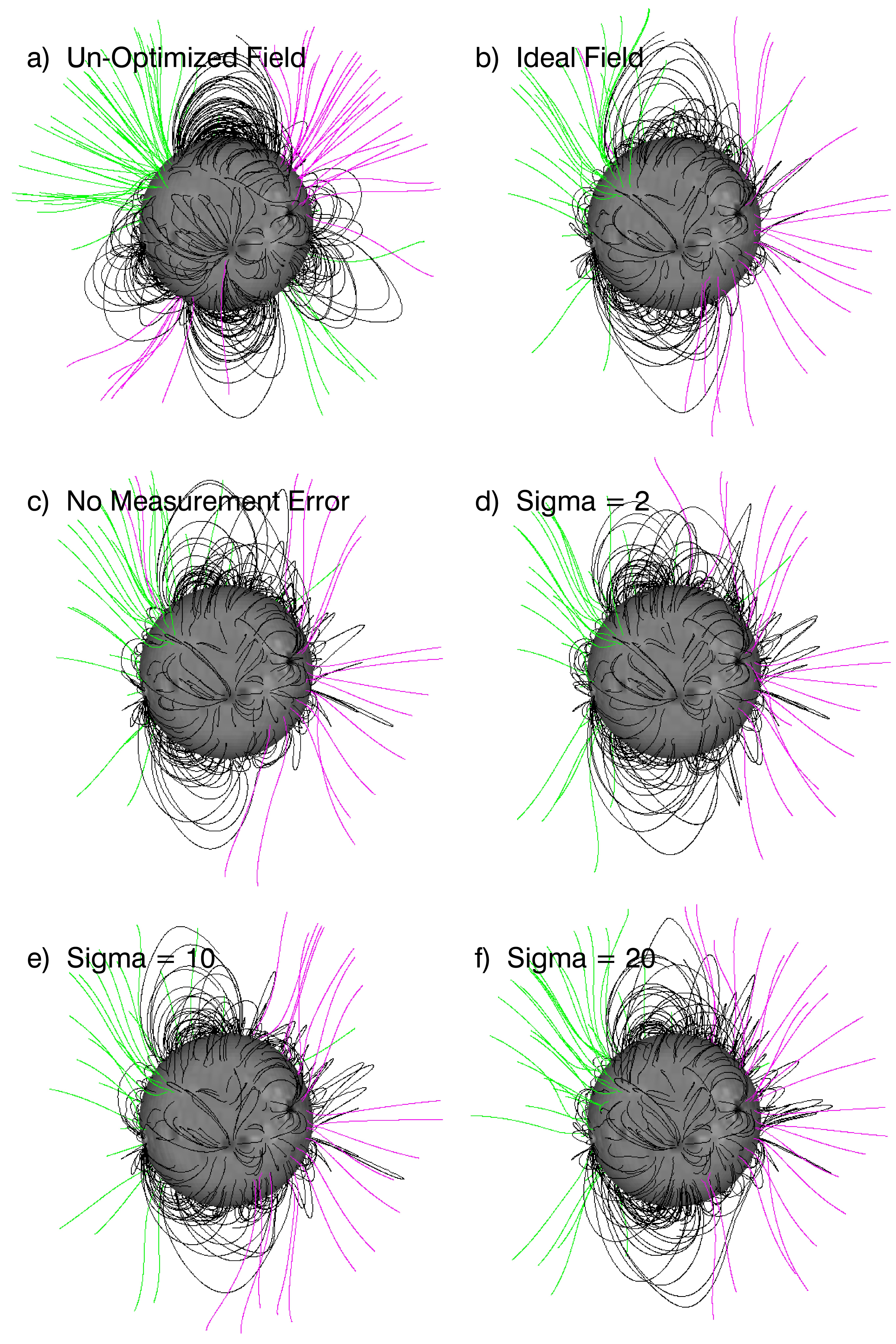}
 \caption{a) Perturbed PFSS field generated from GONG magnetogram from January 10, 2014.  This is the starting point of the model field.  b) Ideal model field from which optimization constraints were drawn.  c) Result of optimization based on ideal constraints with no error in the measured orientation angles.  d)-f)  The model field after optimization with constraints with normally distributed error with standard deviation of $\sigma_{\theta}= 2\degree$, $10\degree$, and $20\degree$.}
\label{fig:noise_hairyballs}
\end{figure}

Figure \ref{fig:noise_hairyballs} shows the ideal field, perturbed field, and several optimized fields corresponding to constraints with different error levels.  Closed magnetic field lines are shown in black, while open field lines are shown in green (for field lines emerging from the photosphere) or magenta (for field lines oriented into the photosphere).  From this figure we can see that the optimized field is substantially closer to the ideal field in all cases.  Some points of comparison highlighting the improvement made using the optimization software include: 1) The ideal field (b) contains a region of magenta colored open field in the bottom right quadrant.  In the un-optimized field (a), this area contains only closed field.  After optimization, however, every field model includes open field in this quadrant.  2) In the un-optimized field there is a large region of green open flux visible in the bottom right quadrant emerging from the far side of the Sun.  In every optimized field this has been eliminated.  3) In the bottom left quadrant of the un-optimized field, a stripe of magenta-colored open flux extends from the center of the disk to the bottom left limb, with a region of large closed loops the north of it.  In the ideal field and in every optimized field, there is no stripe of open flux, and the closed-field region to the north is much more compact and has substantially different connectivity.
\begin{table}[h]
  \centering
  \begin{tabular}{| c  c  c  c  c |}
    \hline
     & $C_{vec}$ & $C_{cs}$ & $1-E_n$ & $1-E_m$  \\
    \hline
    Before & & & & \\
    Optimization & 0.984 & 0.489 & 0.469 & -1.46 \\
    \hline
    $\sigma_{\theta}=0.0\degree$ & 0.999 & 0.936 & 0.892 & 0.644\\
    $\sigma_{\theta}=2.0\degree$ & 0.998 & 0.942 & 0.854 & 0.457 \\
    $\sigma_{\theta}=5.0\degree$  & 0.998 & 0.939 & 0.854 & 0.417 \\
    $\sigma_{\theta}=10.0\degree$  & 0.998 & 0.928 & 0.837 & 0.378 \\
    $\sigma_{\theta}=15.0\degree$  & 0.998 & 0.939 & 0.836 & 0.396 \\
    $\sigma_{\theta}=20.0\degree$  & 0.998 & 0.932 & 0.830 & 0.360 \\
   \hline
    $\sigma_{r}=0.0\degree$ & 0.999 & 0.936 & 0.891 & 0.643 \\
    $\sigma_{r}=2.0\degree$ & 0.997 & 0.944 & 0.815 & 0.334 \\
    $\sigma_{r}=5.0\degree$  & 0.996 & 0.935 & 0.751 & 0.063 \\
    $\sigma_{r}=7.0\degree$  & 0.995 & 0.928 & 0.721 & -0.019 \\
    $\sigma_{r}=10.0\degree$  & 0.993 & 0.923 & 0.681 & -0.117 \\
    $\sigma_{r}=15.0\degree$  & 0.984 & 0.878 & 0.505 & -0.728 \\
    \hline
  \end{tabular}
  \caption{Effect of error in $\theta_{o}(\textbf{r})$ on optimization results.  Here $\sigma_{\theta}$ and $\sigma_{\textbf{r}}$ are the standard deviation of the error in the orientation angles and the position determination, respectively.}
  \label{table:foms}
\end{table}

As in Paper I, we have quantified the improvement in the field after optimization using the figures of merit presented by \citet{schrijver06}.  The vector correlation metric, $C_{vec}$,
\begin{equation}
C_{vec} \equiv \frac{\sum {\mathbf{B}_i \cdot  \mathbf{b}_i}}{( \sum \| \mathbf{B}_i \|^2  \sum \| \mathbf{b}_i \|^2 )^{1/2}} 
\end{equation}
the Cauchy-Schwartz metric,
\begin{equation}
C_{CS} \equiv \frac{1}{M} \sum \frac{\mathbf{B}_i \cdot \mathbf{b}_i}{\| \mathbf{B}_i \| \| \mathbf{b}_i \|} 
\end{equation}
the normalized vector error metric,
\begin{equation}
E_n \equiv \frac{\sum \| \mathbf{b}_i -\mathbf{B}_i \|}{\sum \| \mathbf{B}_i \|}
\end{equation}
and the mean vector error metric
\begin{equation}
E_m \equiv \frac{1}{M} \sum \frac{\| \mathbf{b}_i - \mathbf{B}_i \|}{\|\mathbf{B}_i \|}
\end{equation}
In all cases, $M$ is the number of computation grid cells, $\mathbf{B}_i$ is the ideal magnetic field at point $i$, and $\mathbf{b}_i$ is the approximated magnetic field at point $i$ in our model.  We have not chosen to include the fifth metric used in Paper I because of its limited usefulness in evaluating potential fields.

Table \ref{table:foms} gives the values of these figures of merit for the five \textbf{error} levels evaluated.  (Note that in the table $E_{n}$ and $E_{m}$ have been subtracted from one so that all four columns have ideal values of $1$.)  The range of values chosen for $\sigma_{\theta}$ was meant to cover the largest possible expected error distribution; at $\sigma_{theta} = 20\degree$ the difference between the directions of the local image gradients and the orientation angles should be clearly visible.  As expected, the case with the least error in the constraints produced the closest match to the ideal field, and all of the metrics showed clear improvement over the model before optimization, though $C_{vec}$ (the vector analog to the correlation function) is already very high in the perturbed field.
\subsection{Errors in Localization}
\label{section:los_test}
In order to address the concerns about \ac{los} confusion described in section \ref{section:los}, we repeat the theoretical test again, this time introducing error in the \emph{locations} of the sampled angles.  After generating the locations (radius and latitude) of our set of constraints as described above, we stepped each constraint away from its respective \ac{pos}, along the observer's line of sight.  The angular distances of the sampling points from the \ac{pos} are drawn from a normal distribution with standard deviation $\sigma_{\textbf{r}}$.  Sampling the ideal model field orientations at the locations outside of the \ac{pos}, we fed the resulting $\{\theta_{o}\}$ to the optimization routine, simulating an erroneous assumption that the observed features are in the \ac{pos}.  We repeated this experiment for $\sigma_{\textbf{r}}$ values of $2.0\degree, 5.0\degree, 7.0\degree, 10.0\degree,$ and $15.0\degree$, performing $10$ trials at each value.  We then compared the optimized field for each case to the ideal field using the four figures of merit described in the previous section.
 
The bottom portion of table \ref{table:foms} gives the values of these metrics for each value of $\sigma_{\textbf{r}}$, averaged over ten trials.  The range of $\sigma_{\textbf{r}}$ values was chosen to cover the longitude range over which features might realistically be projected into the \ac{pos} with sufficient contrast to be identified as potential constraints on the field.  At the Carrington rotation rate features should traverse just over $13\degree$ per day, so a standard deviation of $15.0\degree$ covers material within one day's rotation of the \ac{pos}, and it is unusual for the same feature to be distinguishable in \ac{pb} coronagraph images for more than two days.  From the values in table \ref{table:foms} one can see that every \ac{fom} is improved by the optimization, except for the vector correlation in the worst case ($\sigma_{\textbf{r}}=15.0\degree$), which is unchanged from its value before optimization.
\begin{figure}
  \centering
\includegraphics[width=11cm]{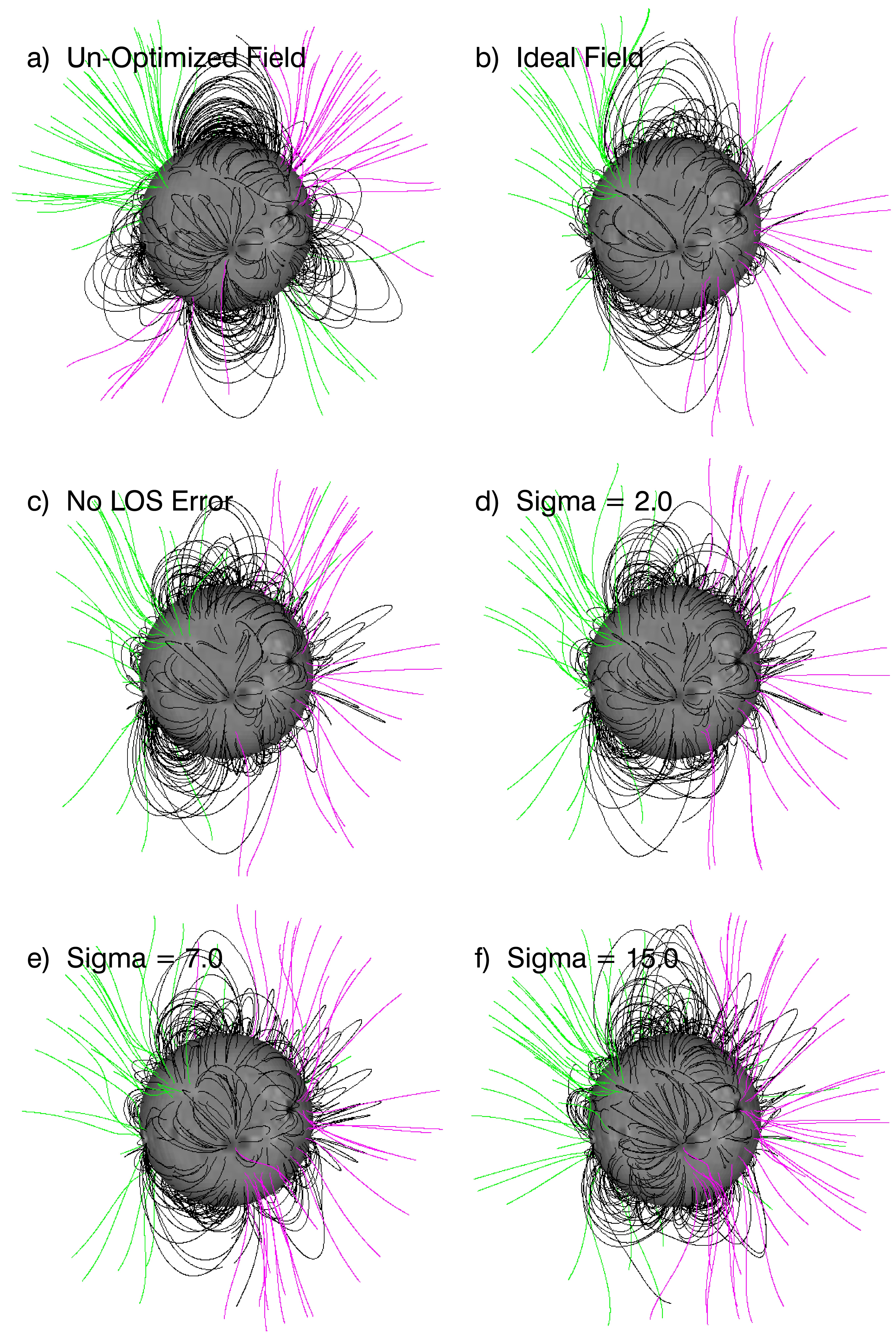}
	  \caption{Magnetic field model sketches.  a) Perturbed PFSS field generated from GONG magnetogram from January 10, 2014.  This is the starting point of the model field.  b) Ideal model field from which optimization constraints were drawn.  c) Result of optimization based on ideal constraints drawn from the POS.  d)-f) Optimized model fields for cases where initial constraints are scattered away from the \ac{pos} along the \ac{los} according to a normal distribution with standard deviations of $\sigma_{\textbf{r}} = 2.0\degree$, $7.0\degree$, and $15.0\degree$, respectively.  Models c)-f) all represent substantial improvement over the perturbed field in a).}
\label{fig:los_hairyballs}
\end{figure}

Figure \ref{fig:los_hairyballs} shows the model magnetic field for several values of $\sigma_{\textbf{r}}$.  The points of comparison discussed in section \ref{section:noisy_constraints} can all be seen to hold true here as well; the open-field regions have been restored to their proper positions after optimization.
\section{Conclusions}
In this paper we have demonstrated the usefulness of coronagraph images for constraining coronal magnetic field models.  We have briefly reviewed the method presented in Paper I for using image-based information to optimize global coronal magnetic field models, and have presented extensions of the theoretical testing published therein, incorporating the kinds of constraints that could be obtained from high-quality coronagraph images.  These new test constraints were distributed spatially primarily in the region of the streamer belt and clustered near the edge of a hypothetical occulter, and included measurement error in the observed orientation angles.  Based on visual inspection of the field models in figure \ref{fig:noise_hairyballs} and the numerical figures of merit in table \ref{table:foms}, the optimization algorithm is able to effect substantial improvement in the field model under these conditions.

We also explored the effects of possible errors in the location of the constraints due to our assumption that the observed features are located in the \ac{pos}.  We presented tests in which the hypothetical image-based constraints revealed the field at locations scattered around the plane of the sky, rather than in it as the optimization software assumes.    The results of the optimization, seen in figure \ref{fig:los_hairyballs} and table \ref{table:foms}, showed that even for constraints scattered as much as $15\degree$ from the \ac{pos}, we're able to measurably improve the magnetic field model from its pre-optimization condition.

\bibliography{master_extrapolation_practicalities}
\end{document}